# Probing FeSi, a *d*-electron topological Kondo insulator candidate, with magnetic field, pressure, and microwaves

Alexander Breindel[1], Yuhang Deng[1], Camilla M. Moir[1], Yuankan Fang[1], Sheng Ran[1], Hongbo Lou[2], Shubin Li[2,3], Qiaoshi Zeng[2], Lei Shu[4,5], Christian T. Wolowiec[1], Ivan K. Schuller[1], Priscila F. S. Rosa[6], Zachary Fisk[7], John Singleton[6], and M. Brian Maple[1]

[1]Department of Physics, University of California, San Diego, CA 92093, United States of America
[2]Center for High Pressure Science and Technology Advanced Research, Pudong, Shanghai 201203, People's Republic of China
[3]Univ Lyon, Université Claude Bernard Lyon 1, CNRS, Institut Lumière Matière, Villeurbanne, France
[4]State Key Laboratory of Surface Physics, Department of Physics, Fudan University, Shanghai 200433, People's Republic of China
[5]Shanghai Research Center for Quantum Sciences, Shanghai 201315, People's Republic of China
[6]Los Alamos National Laboratory, NM 87545, United States of America
[7]University of California, Irvine, CA 92967, United States of America
Corresponding author: M. B. Maple
Present address of Sheng Ran: Department of Physics, Washington University in Saint Louis, Saint Louis, MO 63130, United States of America

*February 23, 2022*

**Abstract**

Recently, evidence for a conducting surface state below 19 K was reported for the correlated $d$-electron small gap semiconductor FeSi. In the work reported herein, the conducting surface state and the bulk phase of FeSi were probed via electrical resistivity $\rho$ measurements as a function of temperature $T$, magnetic field $B$ to 60 T and pressure $P$ to 7.6 GPa, and by means of a magnetic field modulated microwave spectroscopy (MFMMS) technique. The properties of FeSi were also compared to those of the Kondo insulator $SmB_6$ to address the question of whether FeSi is a $d$-electron analogue of an $f$-electron Kondo insulator and, in addition, a "topological Kondo insulator." The overall behavior of the magnetoresistance $MR$ of FeSi at temperatures above and below the onset temperature $T_S$ = 19 K of the conducting surface state is similar to that of $SmB_6$. The two energy gaps, inferred from the $\rho(T)$ data in the semiconducting regime, increase with pressure up to about 7 GPa, followed by a drop which coincides with a sharp suppression of $T_S$. This behavior is similar to that reported for $SmB_6$, except that the two energy gaps in $SmB_6$ decrease with pressure before dropping abruptly at $T_S$. The MFMMS measurements showed a sharp feature at $T_S \approx 19$ K for FeSi, but no such feature was observed at $T_S \approx 4.5$ K for $SmB_6$. The absence of a feature at $T_S$ for $SmB_6$ may be due to experimental issues and will be the subject of a future investigation.

**Introduction**

Recently, electrical resistivity measurements on high-quality single crystals of the correlated $d$-electron small gap semiconductor FeSi revealed a crossover from semiconducting to metallic behavior below ~19 K [1]. The metallic behavior was attributed to the emergence of a conducting surface state below an onset temperature $T_S \approx 19$ K, suggesting that FeSi could be a topological insulator [1]. The correlated $d$-electron compound FeSi [2, 3, 4] has attracted much interest for more than six decades because of its unusual electrical and magnetic properties (e.g., refs. [5, 6, 7, 8, 9, 10, 11, 12, 13, 14, 4]) which are reminiscent of those of $f$-electron Kondo insulators. This has led to the proposal that FeSi is a $d$-electron counterpart of an $f$-electron Kondo insulator (e.g., refs. [7, 13, 15]).

Originally called "hybridization gap semiconductors" because the energy gap is produced by hybridization of localized $f$- and conduction electron states [16, 17, 18], Kondo insulators comprise a set of small gap semiconducting compounds based on lanthanide and actinide elements such as Ce, Sm, Yb, and U with partially-filled $f$-electron shells and an unstable valence [19, 20, 21, 18]. The members within this group that were first identified are the Sm compounds $SmB_6$ [22, 23] and SmS in its collapsed "gold phase" [24, 25]. In both $SmB_6$ and the "gold phase" of SmS, Sm has an intermediate valence of ~2.7 and a nonmagnetic ground state that was attributed to "valence fluctuations" [26, 27, 28, 29]. This was followed by discoveries of Ce-, Yb-, and U-based Kondo insulators starting with $CeFe_4P_{12}$ [20], $YbB_{12}$ [19], and $UFe_4P_{12}$ [20]. Kondo insulators have also been proposed to be topological insulators on theoretical grounds [15, 30], and mounting evidence during the past several years indicates that one of the original and most extensively investigated Kondo insulators, $SmB_6$, is a "topological Kondo insulator" with a conducting surface state that dominates electrical conductivity below $T_S \approx 4.5$ K [31, 32, 33, 34, 35, 36, 37, 38, 39].

The discovery of a conducting surface state below $T_S \approx 19$ K in FeSi and the similarity of many of its properties with those of $SmB_6$ suggests that FeSi may also be a "topological Kondo insulator" [1]. Topological materials have gained much attention in recent years owing to the possibility that topological superconductors, as well as strong topological insulators in proximity to s-wave superconductors, can act as hosts for Majorana modes, which have applications in quantum computing [40]. If FeSi is indeed a "topological Kondo insulator," it could be an attractive candidate for potential applications in spintronics and quantum computing since it is comprised of elements that are relatively inexpensive and compatible with Si-based electronics.

In this work, we have probed the conducting surface state and bulk properties of FeSi with high magnetic fields up to 60 T, high pressures up to 7.6 GPa, as well as microwaves via the magnetic field modulated microwave spectroscopy (MFMMS) technique [41]. We have compared the properties of FeSi to those of the Kondo insulator $SmB_6$ to address the question of whether FeSi is a $d$-electron analogue of an $f$-electron Kondo insulator and, in addition, a topological Kondo insulator. This has also involved making additional magnetoresistance measurements on SmB$_6$. In the initial work on FeSi, we found that the temperature *T* dependence of the electrical resistance *R*(*T*) of high-quality single crystals exhibits semiconducting behavior above 19 K which can be described by a standard activation model with a two-gap feature at higher temperatures and metallic behavior below 19 K [1]. The observation that the normalized electrical resistance *R*(*T*)/*R*(120 K) below 19 K shows a dependence on the dimensions of the FeSi specimens along with the absence of any features in the specific heat at 19 K, indicated the presence of a conducting surface state [1]. Recent research employing scanning tunneling microscopy on high quality FeSi single crystals supports the existence of surface conductivity of FeSi and further illustrates the similarity of its correlated electron properties to those of SmB$_6$ [42].

**Experimental Details**

The FeSi single and SmB$_6$ single crystals were prepared by means of flux growth methods in tin [1] and aluminum [43, 44] fluxes, respectively. Magnetoresistance measurements were carried out at Los Alamos National High Magnetic Field Laboratory (NHMFL) in pulsed fields up to 60 T at temperatures from 0.7 K to 27 K. Electrical resistance measurements at ambient pressure were performed at the University of California, San Diego (UC San Diego) in a Quantum Design DynaCool Physical Property Measurement System (PPMS) at temperatures down to 1.8 K and magnetic fields up to 9 T. Electrical resistance measurements under pressure were carried out at the Center for High Pressure Science and Technology Advanced Research in Shanghai with a diamond anvil cell (DAC) and at the UC San Diego with a hydrostatic piston cylinder cell (PCC).

A single crystalline FeSi sample (bar shape, 25 μm thick and 80 μm long) was compressed to a maximum pressure of 7.6 GPa in a DAC using diamond anvils, each with a culet size of 300 micrometers. A 300 μm thick T301 stainless steel gasket was pre-indented to 48 μm, and the indent bottom was removed with a laser. An insulating layer comprised of a mixture of cubic boron nitride (cBN) and epoxy was compressed into the indent. The laser was used again to make a new hole in the cBN layer. The hole was then filled with sodium chloride (NaCl) which served as a good pressure transmitting medium. Several ruby spheres were placed within the NaCl pressure transmitting medium, serving as a pressure gauge. A standard four-point method was used to measure the resistance of the sample under pressure with 4 µm thick platinum strips as the electrical leads. An AC resistance bridge was used with an amplitude from 0.01 to 0.1 mA, 22 or 33 Hz excitation current. The DAC containing the sample was inserted into a cryostat capable of varying temperature from 300 K to 2 K and magnetic field from 0 T to 3 T. Pressures were determined at room temperature.

A PCC made of nonmagnetic materials was used for measuring the electrical resistivity of FeSi under hydrostatic pressures up to 2.45 GPa. A piece of single crystalline FeSi was placed in a Teflon capsule filled with a liquid pressure-transmitting medium composed of a mixture of n-pentane and isoamyl alcohol (volume ratio 1:1). Mutual inductance coils embedded within the BeCu clamp body were used to measure the ac magnetic susceptibility of a tin or lead superconducting manometer located inside the sample space from which the superconducting critical temperature and, in turn, the pressure was determined. An LR 700 AC resistance bridge was employed to measure the electrical resistance of the sample. A liquid helium Dewar was used to vary the temperature of the sample between room temperature and 1.5 K by adjusting the height of the pressure clamp above the liquid helium bath and by pumping on the liquid helium bath after the clamp was immersed in the liquid.

Single crystal samples of FeSi and $SmB_6$ were studied by means of the magnetic field modulated microwave spectroscopy (MFMMS) technique [41] to see if it would be possible to detect the onset of the conducting surface states inferred from $\rho(T)$ measurements on these compounds. The MFMMS setup consists of a customized Bruker X-band (9.4 GHz) electron paramagnetic resonance (EPR) apparatus with a microwave power source, a dual-mode cavity resonator, lock-in detector, and a 1 T electromagnet. The MFMMS technique measures the reflected microwave power from a sample as a function of temperature [41]. In general, the absorption of microwave power depends on the surface resistance of a material. For example, when a material undergoes a superconducting transition, the decrease in surface resistance reduces the absorption of microwave power at the surface, resulting in a pronounced peak in the MFMMS signal near the superconducting transition.

For the MFMMS measurements, five rod-like single crystal FeSi samples were placed at the bottom of a thin quartz tube which was then flushed with helium gas and sealed with paraffin film. Individual $SmB_6$ single crystal samples were similarly sealed in quartz tubes. The FeSi and $SmB_6$ samples in the bottom of the quartz tubes were then placed at the center of a cavity resonator where the magnetic field component of the $TE_{102}$ mode is at a maximum. The five needle-like FeSi crystals had lengths ranging from 1.15 to 1.9 mm and diameters ranging from 45 to 60 micrometers. Field cooled (FC) measurements were performed at various DC fields set with the electromagnet while the sample temperature was swept at a rate of 1 K per minute using an Oxford helium flow cryostat and temperature control. The application of an external ac magnetic field of 15 Oe at 100 KHz and the use of a lock-in amplifier provided an enhanced signal and reduction of noise for the detection of reflected microwave power. All MFMMS measurements of both FeSi and $SmB_6$ samples were performed at a microwave power of 1 mW.

## Results and Discussion
### Evidence for a conducting surface state

Evidence for a conducting surface state below $T_S \approx 19$ K in a high-quality flux-grown single crystal of FeSi was reported in reference [1]. In Fig. 3 of reference [1], it was shown that $\rho(T)$ in the metallic region below 19 K decreases systematically with decreasing average radius $r$ of an approximately rod-shaped FeSi single crystal of length $L$ as the surface area $A$ to volume $V$ ratio, $A/V = (2\pi r L)/(\pi r^2 L) = 2/r$ increases. This indicates that the contribution to the overall conductivity of the conducting surface state is larger than that of the bulk small gap semiconducting portion of the crystal. Assuming the conducting surface state is confined to a cylindrical shell of radius $r$ and effective thickness $t$, the normalized electrical resistance $R(T)/R(120 \text{ K})$ in the metallic region below $T_S \approx 19$ K can be expressed as $R(T)/R(120 \text{ K}) = \rho_m(T)(L/2\pi r t)/\rho_s(120 \text{ K})(L/\pi r^2) = [\rho_m(T)/\rho_s(120 \text{ K})](r/2t) \propto r$, where $\rho_m(T)$ is the electrical resistivity in the metallic region below $T_S$, and $\rho_s(120 \text{ K})$ is the electrical resistivity in the semiconducting region at 120 K. Thus, at fixed $T$, one expects the normalized electrical resistance $R(T)/R(120\text{ K})$ to decrease linearly with decreasing $r$ as the sample is thinned and the conducting surface state region contributes more to the conductivity. In the semiconducting region above 19 K, where the surface state contribution is negligible, $R(T)/R(120\text{ K}) = \rho_s(T)(L/\pi r^2)/\rho_s(120\text{ K})(L/\pi r^2) = \rho_s(T)/\rho_s(120\text{ K})$ is independent of $r$ and all of the normalized $R(T)$ curves collapse onto one another. This behavior is illustrated in Fig. 1 in the plot of $R(T)/R(120\text{ K})$ vs. average diameter $D = 2r$ of an approximately cylindrical rod shaped sample of FeSi at 8 different temperatures between 2 K and 30 K (from the data reported in Fig. 3 of reference [1]).

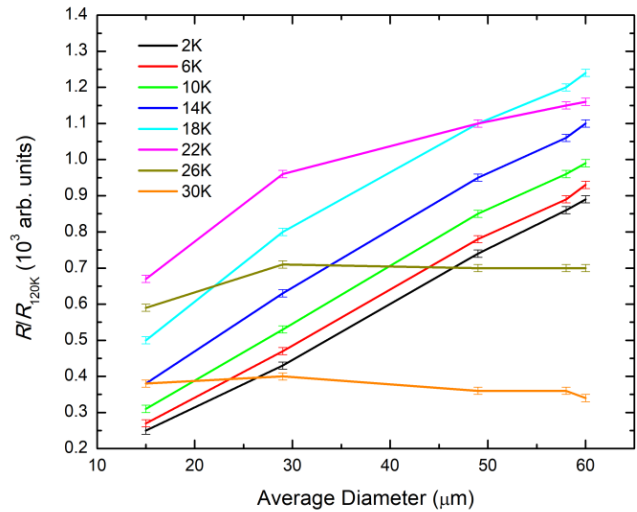

**Figure 1** Electrical resistance $R$, normalized to its value at 120 K, vs. the average diameter of an approximately cylindrical rod-shaped sample of FeSi, from the data reported in [1]. Error bars are included to account for errors in the data retrieval. The linearity of the plots suggests the formation of a conducting surface state below around 19 K, because of the increased surface area to volume ratio associated with thinning the samples [1].

**Magnetoresistance**

Figure 2 shows the transverse electrical resistance $R_\perp$ vs. $B$, measured with the magnetic field $B$ perpendicular to the direction of the current, for a single crystalline FeSi specimen at various temperatures $T$ between 0.7 K and 27 K. The measurements were performed upon field upsweep and downsweep, and the downsweep data were presented because there could be a jump in the resistivity at the beginning of the pulse due to the large $dB/dt$ There is a small positive MR region at low magnetic fields, followed by a negative MR region at higher fields. The transition field $B_m$ between these two regions is defined as the field where the curvature in $R_\perp(T)$ changes from negative to positive. A plot of $B_m$ vs. $T$ displayed in the inset of the $B$–$T$ plot in Fig. 2 shows that $B_m$ decreases with $T$ and bends over and appears to extrapolate to zero at a value of $T$ near $T_S \approx 19$ K, the onset temperature of the conducting surface state [1]; however, within the experimental uncertainty, $B_m$ saturates to value of ~0.5 T at 20 K and 27 K.

The $R_\perp$ vs. $B$ curves also reveal that $R_\perp$ exhibits a maximum as a function of $T$ at $T_S \approx 20$ K, as can be seen more clearly in Fig. 3. This weak dependence of $T_S$ on $B$ is represented by the nearly vertical line in the inset of Fig. 2. The lines representing $B_m(T)$ and $T_S(B)$ in Fig. 2 divide the $B$–$T$ plane into three regions – conducting surface state (CSS), $dR/dT > 0$, positive magnetoresistance $MR > 0$ (except for the two lowest temperatures of 0.7 K and 1.47 K), where $MR \equiv [R(B) - R(0)]/R(0)$ (lower left); CSS, $dR/dT > 0$ with $MR < 0$ (upper right), Kondo insulator state (KIS), $dR/dT < 0$ with $MR < 0$ (far right). Positive $MR$ within the CSS could be due to the metallic character of the CSS, while the negative $MR$ within the CSS could be associated with a decrease in the energy gap of the semiconducting state of bulk FeSi and a suppression of the CSS with increasing $B$.

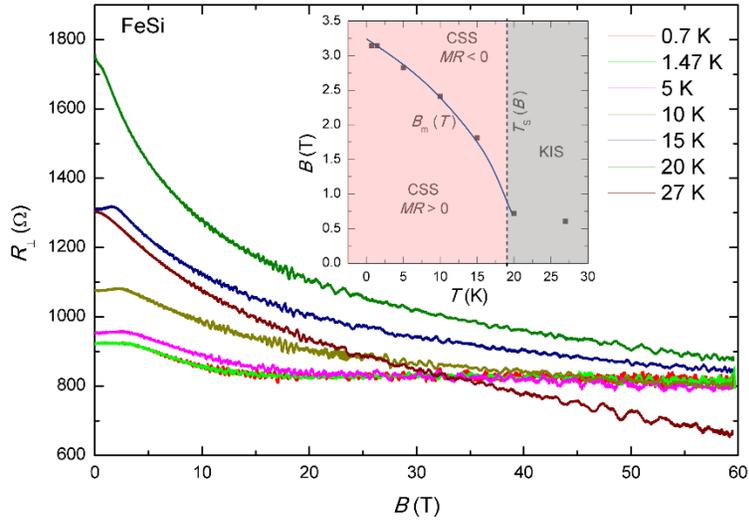

**Figure 2** Transverse electrical resistance $R_\perp$, measured with the magnetic field $B$ perpendicular to the direction of the current, vs. $B$ at various temperatures $T$ between 0.7 K and 27 K. The data presented here were from field downsweep. There is a small positive MR region at low magnetic fields, followed by a negative MR region at higher fields. The transition field $B_m$ between these two regions is defined as the field where the curvature in $R_\perp(T)$ changes from negative to positive. Inset: $B$–$T$ plot in which the curves $B_m(T)$ and $T_S(B)$ divide the $B$–$T$ plane into three regions – CSS, $dR/dT > 0$ with $MR > 0$ (lower left); CSS, $dR/dT > 0$ with $MR < 0$ (upper right); KIS, $dR/dT < 0$ with $MR < 0$ (far right). The meaning of the symbols in the inset of Fig. 2 are as follows: CSS – conducting surface state, KIS – Kondo insulator state, and $MR \equiv [R(B) - R(0)]/R(0)$ – magnetoresistance.

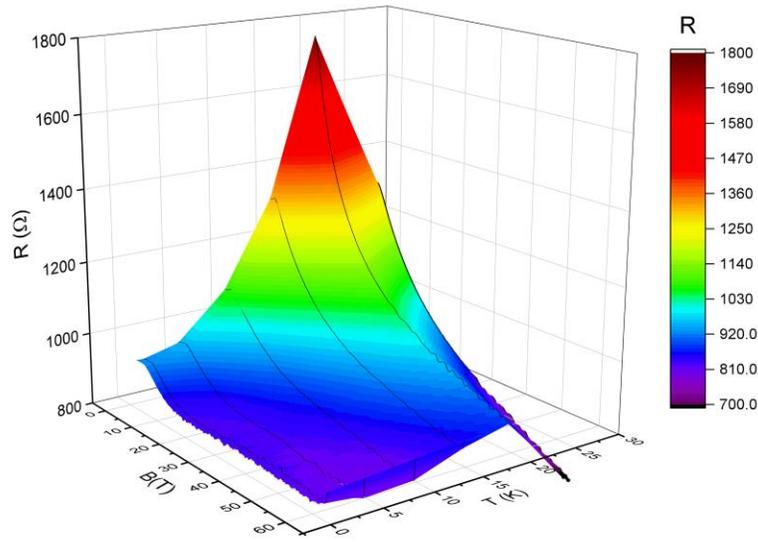

**Figure 3** 3D surface plot derived from the transverse resistance $R_\perp$ vs. magnetic field $B$ and temperature $T$ data shown in Figure 2. There is a clear peak in the $R_\perp(T)$ data around 20 K for all fields, associated with the transition to the conducting surface state [1], as well as a peak in the $R_\perp(B)$ data associated with the change in *MR* from positive for the conducting surface state to negative for the semiconducting bulk KIS state.

The transverse resistance $R_\perp$ is compared with the longitudinal resistance $R_\parallel$ vs. *B*, measured with the magnetic field *B* parallel to the direction of the current, vs. *B* at two temperatures below $T_S$, 0.7 K and 10 K, in Fig. 4. A schematic of the geometry of the transverse $R_\perp$ and longitudinal $R_\parallel$ resistance measurements, where the magnetic field is oriented perpendicular and parallel to the longitudinal axis of the sample, respectively, is shown in the inset of Fig. 4. It is clear from the data in Fig. 4 that there is significant anisotropy in the resistance *R* vs. *B* associated with the surface state, despite the cubic crystal structure of FeSi [1]. These results are consistent with our previously published magnetoresistance data on FeSi samples, and, as explained in our previous work, the difference caused by the change in field orientation can be understood as a consequence of a positive field response of the contribution of the surface conductivity to the overall magnetoresistance [1].

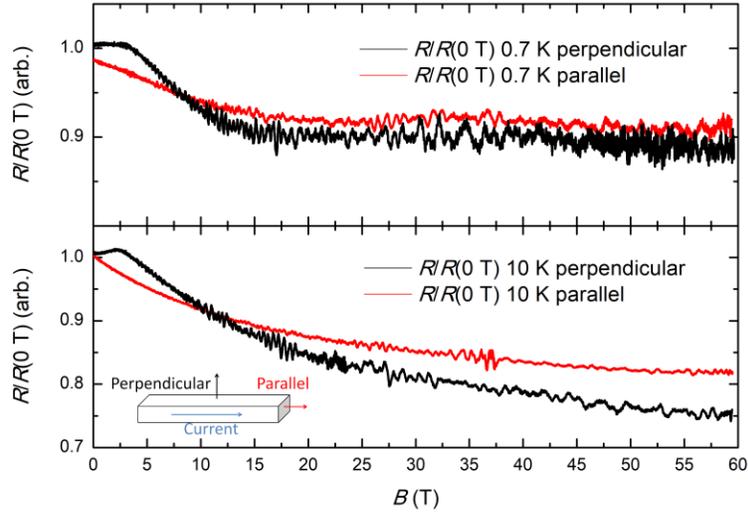

**Figure 4** Anisotropic electrical resistance $R(B)$, normalized to the value of $R$ at $B = 0$ T, $R/R(0\ T)$, vs. magnetic field $B$ curves at 0.7 K (top) and 10 K (bottom). The geometries used in the $R(B)/R(0\ T)$ measurements are shown in the inset. The current is directed along the longitudinal axis of the FeSi single crystal. The transverse resistance $R_\perp$ (black curves) and longitudinal resistance $R_\parallel$ (red curves) were measured with the magnetic field perpendicular and parallel to the longitudinal axis of the FeSi single crystal, respectively. The longitudinal axis of the FeSi single crystal corresponds to the [111] direction of the cubic crystal structure.

**Comparison of magnetoresistance measurements on FeSi and SmB$_6$**

Magnetoresistance measurements performed on SmB$_6$ samples show evidence of a crossover at about 4.5 K, as can been in Fig. 5 (b), similar to those shown in Chen *et al.* [37]. Based on the data for FeSi from Fang *et al.* [1], we see the same minimum in the magnetoresistance (see Fig. 5 (a)). The field dependent data for the magnetoresistance *MR* at various temperatures are shown for SmB$_6$ in Fig. 5 (d) taken from Chen *et al.* [37] and for FeSi in Fig. 5 (c), based on the data shown in Fig. 2. In both FeSi and SmB$_6$, there is a large temperature dependence of the *MR* (gray and pink regions) in Figs. 5 (a), (b) around their respective conducting surface state onset temperatures $T_S$. Similar behavior for FeSi and SmB$_6$ can also be seen in the field dependence of the *MR* where below (above) $T_S$, there is a weak (strong) dependence of *MR* on field.

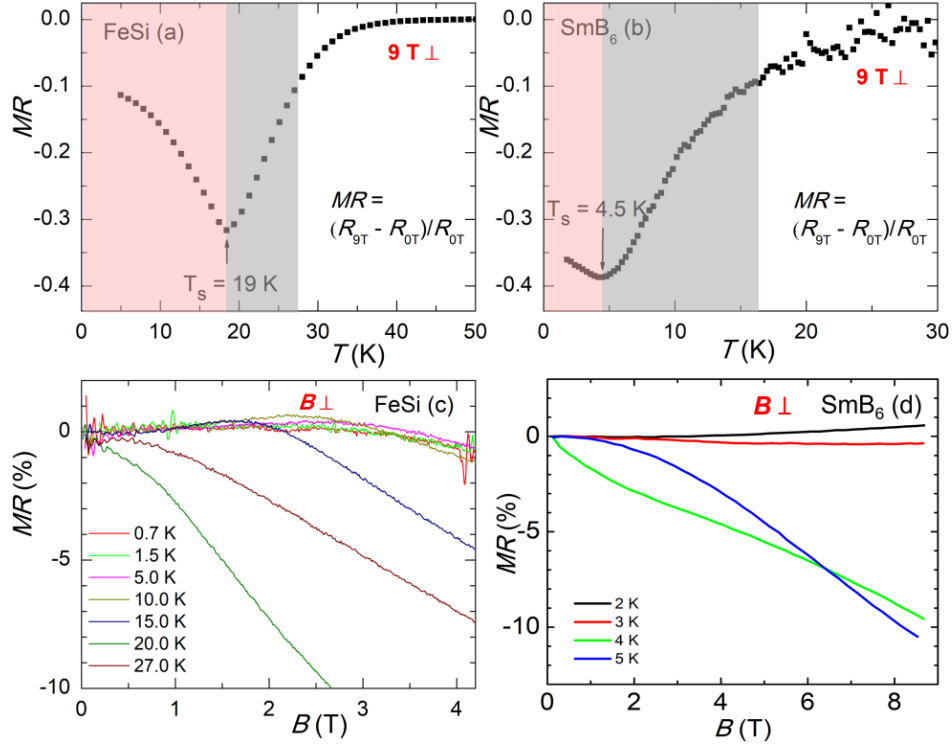

**Figure 5.** Comparison of the magnetoresistance MR of FeSi and SmB$_6$ as a function of temperature T through $T_S$ at a magnetic field B = 9 T [panels (a) and (b)] and as a function of B at various values of T below and above $T_S$ [panels (c) and (d)]. The magnetoresistance is defined as $MR = [R(B) - R(0\ \text{T})]/R(0\ \text{T})$, where B is the applied magnetic field. Data in panel (a) are from Fang *et al.* [1], in panel (b) from this work, in panel (c) from this work (data in Fig. 2), and in panel (d) from Chen *et al.* [37].

**High pressure studies and T vs. P phase diagram**

Shown in Fig. 6 are the T-dependent resistance R(T) curves for FeSi from 2 K to room temperature at various pressures up to 7.6 GPa. There is a persistent peak in the resistance around 20 K for all pressures except 7.6 GPa signifying the striking change in the transport behavior of FeSi at $T_S$. On the higher temperature side of the peak in R(T), the FeSi sample exhibits semiconducting-like behavior with $\frac{dR}{dT} < 0$, whereas on the lower temperature side, the sample displays metallic behavior ($\frac{dR}{dT} > 0$). The same phenomenon has been observed at ambient pressure [1] and has been attributed to the emergence of the conducting surface state in FeSi below the temperature $T_S$ at which the peak in R(T) occurs. As the pressure is increased to 7.6 GPa, the well-defined peak in $R(T)$ vanishes and the resistance of the sample continues to increase as $T$ is reduced to a base temperature of 2 K. This suggests the occurrence of a pressure-induced transition from metallic to insulating-like behavior of the surface state at this pressure. The B20 crystal structure of FeSi (ϵ-FeSi) is quite stable under high pressure. At room temperature, no phase transition was observed to at least 36 GPa. Even with laser heating to above 1000 K, the high-pressure B2 phase (CsCl type FeSi) persists to pressures at least above 14 GPa [45, 46]. Consequently, the pressure-induced change in $R(T)$ of FeSi at 7.6 GPa could be ascribed to an electronic phase transition, and independent of any structural transition.

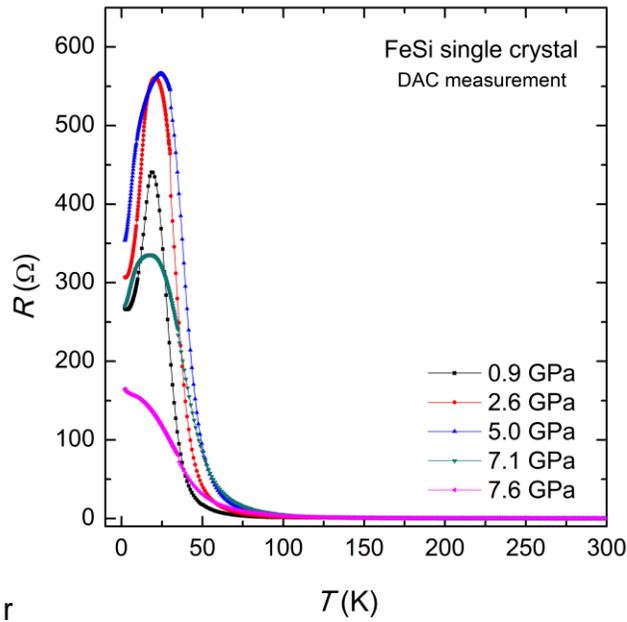

**Figure 6** Electrical resistance $R$ vs. temperature $T$ for an FeSi single crystal at various pressures up to 7.6 GPa measured in DAC experiments.

As noted above, in many reports [47, 21, 7, 13, 15] it has been proposed that FeSi is a $d$-electron Kondo insulator. For a Kondo insulator, a narrow hybridization gap [16, 18] (Kondo gap) develops below a characteristic temperature called the Kondo temperature ($T_K$), because of the coherent spin-dependent scattering of itinerant electrons by the lattice of $d$- or $f$-electron localized magnetic moments. This Kondo scenario is partly consistent with what we found for the $R(T)$ behavior of FeSi. As shown in Fig. 7, $T_K$ is defined as the temperature below which the resistance can be described by a gapped semiconducting activation model. The energy gap $\Delta$ has been extracted from an Arrhenius law,

$$R = R_o \exp(\Delta/2k_B T) \tag{1}$$

where $R_o$ is a constant. The energy gap $\Delta$ can be taken from the slope of the linear portion of a plot of $\ln R$ vs $1/T$. This is illustrated in the $\ln R$ vs. $1/T$ plot in Fig. 7 based on measurements of $R(T)$ for FeSi at 0.9 GPa. The linear region from 70 K to 160 K ($T_K$) corresponds to an energy gap $\Delta_1$ = 57 meV, while the linear region from 37 K to 57 K yields an energy gap $\Delta_2$ = 36 meV. Thus, at 0.9 GPa, $R(T)$ of FeSi evolves from a non-activated regime at $T_K$ = 160 K to an activated regime characterized by an energy gap $\Delta_1$ = 57 meV, then to another activated regime characterized by a smaller energy gap $\Delta_2$ = 36 meV, and finally to a regime involving a conducting surface state below $T_S$ = 19 K.

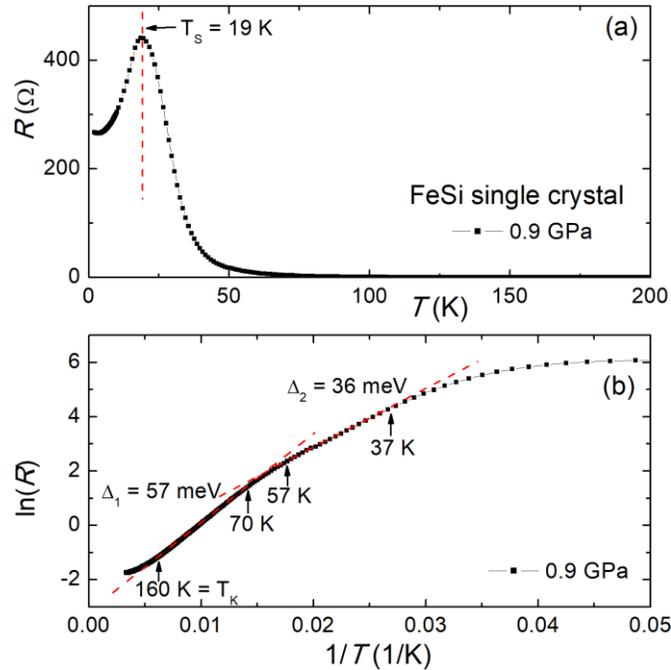

**Figure 7** (a) A representative plot of $R(T)$ at $P$ = 0.9 GPa illustrating the location of $T_S$ at the peak in $R(T)$ marked by the vertical dashed red line. (b) A plot of $\ln R$ vs. $1/T$ that allows for the extraction of the the energy gaps $\Delta_1$ (in the temperature range $T_K$ = 160 to 70 K) and $\Delta_2$ (in the range $T$ = 57 to 37 K) according to an Arrhenius law. Within each region, $R(T)$ is fitted with an Arrhenius law (See Equation (1) and discussion in text). The same procedure was used to determine the two energy gaps $\Delta_1$ and $\Delta_2$ at different pressures based on $R(T)$ measurements performed in PCC and DAC experiments.

The pressure dependences of both energy gaps are plotted in Fig. 8 (a). The energy gaps initially increase with pressure and then begin to decrease around 7 GPa. Correspondingly, in Fig. 6 it can be seen that the peak resistance also first increases and then decreases with increasing pressure, a direct result from the change in the energy gaps under pressure. Furthermore, there is a correspondence of the suppression of the energy gap and the drop in $T_S$, indicated by the vertical dashed line in Fig. 8. The closing of the energy gaps is correlated with the disappearance of the conducting surface state in FeSi. The pressure dependence of the Kondo temperature and the temperature $T_S$ of the onset of the conducting surface state $T_K(P)$ and $T_S(P)$ for the single crystalline FeSi sample are plotted in a $T$ vs. $P$ phase diagram in Fig. 8 (b). Above $T_K$, the sample can be characterized as a bad metal [37] in which itinerant electrons are incoherently scattered by the $d$-electron localized magnetic moments, yielding a very high resistivity compared to that of a simple metal. Between $T_K$ and $T_S$, FeSi is expected to be a Kondo insulator, whereas below $T_S$, a conducting surface state appears, as discussed in our earlier publication [1].

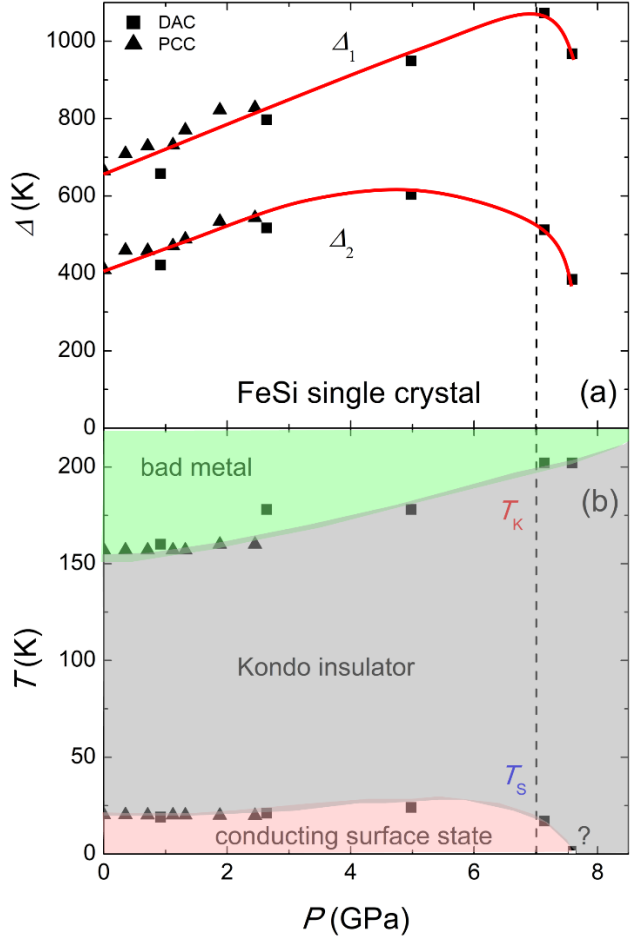

**Figure 8** (a) Evolution of the two energy gaps $\Delta_1$ and $\Delta_2$ with pressure obtained from electrical resistivity measurements in PCC and DAC experiments. (b) $T$-$P$ phase diagram for the FeSi single crystal. The red lines in (a) and the boundaries between different phases are guides to the eye. The question mark in the lower right of the panel indicates the estimation of $T_S$ at 7.6 GPa.

Since the proposal of a topological Kondo insulator in 2010 [15], the $f$-electron Kondo insulator $SmB_6$ was considered a prime candidate. In addition to the many common attributes shared by FeSi and $SmB_6$, we have found that these two compounds display similar electrical transport properties at both ambient and high pressure, providing further evidence that FeSi is a $d$-electron analog of the $f$-electron Kondo insulator $SmB_6$. At ambient pressure, Chen *et al.* also observed two-gap-semiconducting behavior in high quality $SmB_6$ at intermediate temperatures and a conducting surface state below 5 K [37].

At high pressures, this study reveals that the energy gaps of FeSi increase with pressure before they start to fall upon increasing pressure above 7 GPa. Early in 1997, Bauer *et al.* [48] also reported an increase of the energy gap of FeSi under pressures up to 1.3 GPa, with no indication of metallization even at 9.4 GPa. The reason for this discrepancy is not known, but could be due to differences in sample quality or distribution of pressure between the two experiments. Extending the pressure range for FeSi is our future goal to determine the pressure at which the energy gaps vanish. The closure of the Kondo gap at similar critical pressures (4 – 7 GPa), accompanied by a fundamental change in $R(T)$ for $SmB_6$ has been reported by several groups [49, 50, 51, 52, 53]. Using the Anderson lattice model for a Kondo insulator, the indirect hybridization gap obtained by fitting the activation behavior of $R(T)$ is of the order of $V^2/D$, where $V$ is the hybridization energy between the localized *d-* or *f-* and conduction electron states and $D$ the half bandwidth of the conduction band. Both $V$ and $D$ are expected to increase asymptotically with decreasing inter-atomic distance with increasing pressure. This asymptotic behavior is complicated by the node in the Si 3*p* radial wave function and the angular dependence of the atomic wave functions. A non-monotonic pressure dependence of the Kondo gap may appear because of this complication or the periodic Anderson lattice model is not applicable to this situation. At sufficiently high pressure, the hybridization and the hybridization gap would finally vanish, supported by the observations of the gap closure in $SmB_6$ and $Ce_3Bi_4Pt_3$ [54, 55], two typical Kondo insulators. For FeSi, the scenario is even more complicated because the 3*d* state is less localized than the 4*f* state. It would be interesting to explore whether the gaps in FeSi under pressure would finally collapse, since this might clarify the validity of the hybridization model for this *d*-electron counterpart of a Kondo insulator.

**Magnetic Field Modulated Microwave Spectroscopy (MFMMS)**

To gain further information about the transition to a conducting surface state at 19 K in FeSi, magnetic field modulated microwave spectroscopy (MFMMS) measurements were performed on single crystals of FeSi as a function of temperature from 4 K to 100 K in both zero field as well as in an applied DC magnetic field of 500 Oe, the results of which are shown in Fig. 9. In zero field, there are two sharp peaks with an onset of the upper peak at approximately 19 K (see Fig. 9 (a)). At an applied DC magnetic field of 500 Oe, the main peak in MFMMS intensity has shifted to higher temperature with an onset at $T$ = 21 K (see Fig. 9 (b)).

The peaks in the MFMMS signal observed for single crystalline FeSi are reminiscent of the large peaks observed at the onset of a superconducting transition, where the decrease in the surface resistivity during the superconducting transition is the cause for the spike in microwave absorption. The MFMMS signal for the FeSi samples is somewhat unexpected and remarkable for its correspondence to the insulating to metallic transition to the surface state at $T_S$ = 19 K. Previous MFMMS experiments on vanadium sesquioxide ($V_2O_3$), which exhibits a metal to insulator transition with a six order of magnitude change in electrical resistance upon cooling at 160 K, show no peak-like signature associated with the resistivity change [41]. Similar signals have been observed in $PrSi_2$ which has a ferromagnetic transition at 11 K. In this case the MFMMS signal appears with a negative phase [41]. The only other material which showed a peak similar to that of a superconducting transition was a single crystal sample of GaMnAs, which exhibits an insulator-metal transition coincident with a paramagnetic to ferromagnetic transition [41]. Here, the observation of the positive peak-like signal is very unique in the FeSi system presented in this paper.

As a comparison, MFMMS measurements were also performed on two $SmB_6$ samples in zero applied DC field as shown in Fig. 9 (b). Unlike FeSi, the $SmB_6$ MFMMS signal contained no signatures of any transition for 3.8 K < $T$ < 100 K. It is possible that the low-temperature limit of the experiment was not low enough to access the transition to the conducting surface state, reported to be at $T_s$ ≈ 4.5 for $SmB_6$, according to resistivity measurements. The absence of any signal in the MFMMS measurement of $SmB_6$ may be due to the condition of the surface of the $SmB_6$ sample that was not of sufficiently high quality or that there are differences in the underlying physics between the two compounds.

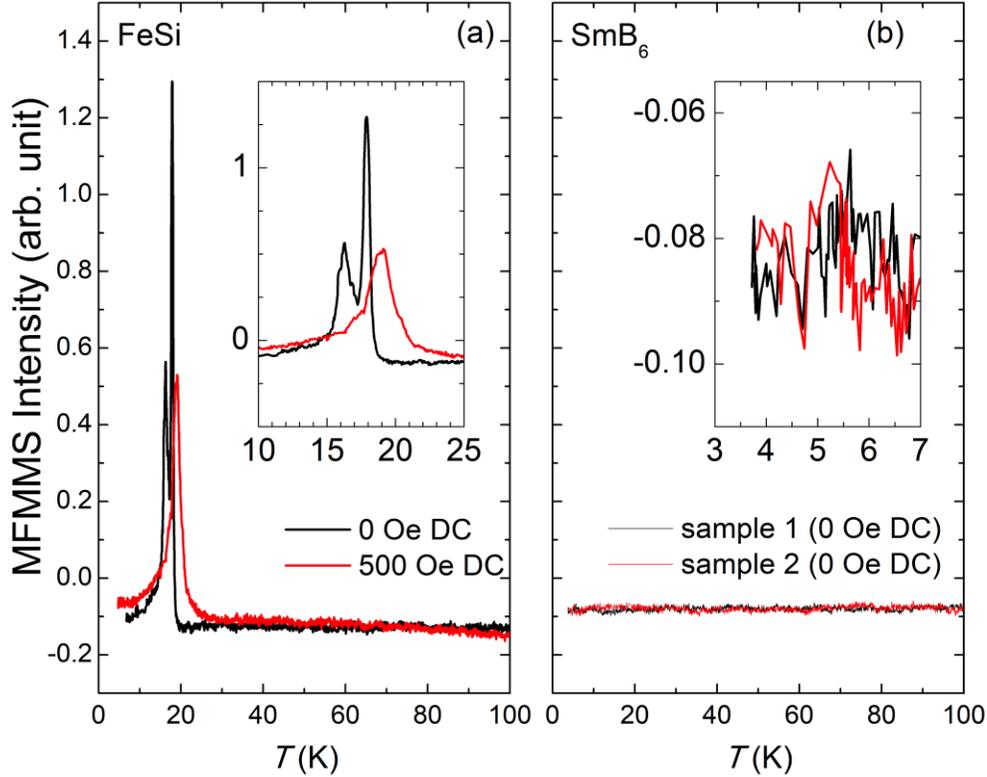

**Figure 9.** Temperature dependence of the microwave absorption signal (MFMMS) intensity) for both FeSi and SmB$_6$. (a) FeSi: The MFMMS signal for FeSi at 0 Oe (black) and 500 Oe (red) applied DC magnetic field. The intensity decreases with field and the onset temperature increases from 19 K at 0 Oe to 21 K at 500 Oe. (b) SmB$_6$: The MFMMS signal for two different SmB$_6$ samples in zero applied DC field. No peaks were observed for repeated measurements down to 4K.

**Concluding remarks**

The conducting surface state and the bulk phase of FeSi were probed via electrical resistivity $\rho$ measurements as a function of temperature $T$, magnetic field $B$ to 60 T and pressure $P$ to 7.6 GPa, and by means of a modulated microwave spectroscopy (MFMMS) technique. The linear plots of $R(T)/R(120\ K)$ vs. the average diameter of an approximately cylindrical rod-shaped sample of FeSi based on the data reported in [1] are consistent with the formation of a conducting surface state below 19 K. The anisotropy of the transverse and longitudinal $MR$ also supports the existence of a surface state below $T_S$. The overall behavior of the $MR$ of FeSi at temperatures above and below the onset temperature $T_S$ = 19 K of the conducting surface state is similar to that of SmB$_6$. The behavior of $R(T)$ of FeSi under pressure reveals that the two energy gaps, inferred from the $R(T)$ data in the semiconducting regime, increase with pressure up to about 7 GPa, followed by a drop which coincides with a sharp suppression of the onset temperature $T_S$ of the conducting surface state. Experiments on $SmB_6$ yielded a similar collapse of the surface state upon closing the two energy gaps with pressure, but with a decrease in the magnitude of the energy gaps with pressure prior to their collapse [52, 53]. The MFMMS measurements showed a sharp feature at the onset of the surface state at $T_S \approx 19$ K for FeSi, but no such feature was observed at $T_S \approx 4.5$ K for $SmB_6$. However, the absence of a feature at $T_S$ for $SmB_6$ should not be regarded as conclusive for the reasons noted above and will be the subject of a future investigation. Although there are competing interpretations [56, 57], most researchers regard the Kondo insulator $SmB_6$ as a topological material [30, 38]. The similarity of the electrical and magneto-transport properties of FeSi and SmB6 found in this work suggests that FeSi is a Kondo insulator, with evidence for the existence of a conducting surface state and topological behavior. This study may extend the scope of topological Kondo insulators from $f$- to $d$-electron materials.

**Acknowledgements**

A. B., Y. D., C. M., Y. F., and M. B. M. would like to acknowledge funding from DOE BES Grant No. DE FG02-04-ER46105 (*materials synthesis*, *high pressure measurements*) and NSF/DMR-1810310 (*physical properties measurements*). P. F. S. R. and Z. F. acknowledge support from the Los Alamos Laboratory Directed Research and Development program (*$SmB_6$ synthesis*). Q. Z. acknowledges the financial support from NSFC (No. 51871054) (*high pressure transport with DACs*). The research performed by L. S. was supported by the Shanghai Municipal Science and Technology (Major Project Grant Nos. 2019SHZDZX01 and 20ZR1405300). J. S. acknowledges funding from NSF/DMR-1157490/1644779 and the State of Florida (*high field transport*), and funding from the DOE BES program "Science in 100 T" that enabled the development and construction of the probes and electronics used in the pulsed-field measurements. C. T. W and I. S. acknowledge funding from DOE BES grant No. DE FG02-87-ER45332 (*MFMMS*). The authors would like to thank Prof. P. Riseborough for discussions about Kondo insulators under pressure.

**References**

[1] Y. Fang, S. Ran, W. Xie, S. Wang, Y. S. Meng, and M. B. Maple. "Evidence for a conducting surface ground state in high-quality single crystalline FeSi," *Proc. Natl. Acad. Sci. U.S.A.* **115**, 8558 (2018). DOI: https://doi.org/10.1073/pnas.1806910115

[2] C. Fu and S. Doniach, "Model for a strongly correlated insulator: FeSi," *Phys. Rev. B* **51**, 17439 (1995).

[3] J. M. Tomczak, K. Haule, and G. Kotliar, "Signatures of electronic correlations in iron silicide," *Proc. Natl. Acad. Sci. U.S.A.* **109**, 3243 (2012). https://doi.org/10.1073/pnas.1118371109

[4] M. Arita, K. Shimada, Y. Takeda, M. Nakatake, H. Namatame, M. Taniguchi, H. Negishi, T. Oguchi, T. Saitoh, A. Fujimori, and T. Kanomata, "Angle-resolved photoemission study of the strongly correlated semiconductor FeSi," *Phys. Rev. B* **77**, 205117 (2008). DOI: 10.1103/PhysRevB.77.205117

[5] H. Watanabe, K. Ito, and H. Yamamoto, "Neutron diffraction study of intermetallic compound FeSi," *J. Phys. Soc. Jpn.* **18**, 995 (1963).

[6] V. Jaccarino, G. K. Wertheim, J. H. Wernick, L. R. Walker, and S. Arajs, "Paramagnetic excited state of FeSi," *Phys. Rev.* **160**, 476 (1967).

[7] Z. Schlesinger, Z. Fisk, Hai-Tao Zhang, M. B. Maple, J. DiTusa, and G. Aeppli, "Unconventional charge gap formation in FeSi," *Phys. Rev. Lett.* **71**, 1748 (1993). DOI: https://doi.org/10.1103/PhysRevLett.71.1748

[8] B. C. Sales, E. C. Jones, B. C. Chakoumakos, J. A. Fernandez-Baca, H. E. Harmon, J. W. Sharp, and E. H. Volckmann, "Magnetic, transport, and structural properties of Fe$_{1-x}$Ir$_x$Si," *Phys. Rev. B* **50**, 8207 (1994).

[9] D. Mandrus, J. L. Sarrao, A. Migliori, J. D. Thompson, and Z. Fisk, "Thermodynamics of FeSi," *Phys. Rev. B* **51**, 4763 (1995).

[10] A. Damascelli, K. Schulte, D. van der Marel, and A. A. Menovsky, "Infrared spectroscopic study of phonons coupled to charge excitations in FeSi," *Phys. Rev. B* **55**, R4863 (1997).

[11] S. Paschen, E. Felder, M. A. Chernikov, L. Degiorgi, H. Schwer, H. R. Ott, D. P. Young, J. L. Sarrao, and Z. Fisk, "Low-temperature transport, thermodynamic, and optical properties of FeSi," *Phys. Rev. B* **56**, 12916 (1997).

[12] K. Breuer, S. Messerli, D. Purdie, M. Garnier, M. Hengsberger, Y. Baer, and M. Mihalik, "Observation of a gap opening in FeSi with photoelectron spectroscopy," *Phys. Rev. B* **56**, R7061 (1997).

[13] J. F. DiTusa, K. Friemelt, E. Bucher, G. Aeppli, A. P. Ramirez, "Metal-insulator transitions in the Kondo insulator FeSi and classic semiconductors are similar," *Phys. Rev. Lett.* **78**, 2831 (1997).

[14] K. Ishizaka, T. Kiss, T. Shimojima, T. Yokoya, T. Togashi, S. Watanabe, C. Q. Zhang, C. T. Chen, Y. Onose, Y. Tokura, and S. Shin, "Ultraviolet laser photoemission spectroscopy of FeSi: Observation of a gap opening in density of states," *Phys. Rev. B* **72**, 233202 (2005). DOI: 10.1103/PhysRevB.72.233202

[15] M. Dzero, K. Sun, V. Galitski, and P. Coleman, "Topological Kondo insulators," *Phys. Rev. Lett.* **104**, 106408 (2010).
DOI: https://doi.org/10.1103/PhysRevLett.104.106408

[16] N. F. Mott, "Rare-earth compounds with mixed valencies," *Phil. Mag.* **30**, 403 (1974). DOI: https://doi.org/10.1080/14786439808206566

[17] P. Coleman, "Heavy Fermions: Electrons at the Edge of Magnetism," *Handbook of Magnetism and Advanced Magnetic Materials,* Helmut Krönmuller and Stuart Parkin, eds. (John Wiley and Sons, Ltd., New York, 2007), Vol. 1, pp 95–148.

[18] P. S. Riseborough, "Heavy fermion semiconductors," *Adv. Phys.* **49**, 257 (2000). DOI: 10.1080/000187300243345

[19] M. Kasaya, F. Iga, K. Negishi, S. Nakai, and T. Kasuya, "A new and typical valence fluctuating system, YbB$_{12}$," *J. Magn. Magn. Matter* **31**, 437 (1983).

[20] G. P. Meisner, M. S. Torikachvili, K. N. Yang, and M. B. Maple, "UFe$_4$P$_{12}$ and CeFe$_4$P$_{12}$: Nonmetallic isotypes of superconducting LaFe$_4$P$_{12}$," *J. Appl. Phys.* **57**, 3073 (1985). http://dx.doi.org/10.1063/1.335217

[21] G. Aeppli and Z. Fisk, "Kondo insulators," *Comments Cond. Mat. Phys.* **16**,155 (1992).

[22] A. Menth, E. Buehler, and T.H. Geballe, "Magnetic and semiconducting properties of SmB$_6$," *Phys. Rev. Lett.* **22**, 295 (1969).

[23] J. C. Nickerson, R. M. White, K. N. Lee, R. Bachmann, T. H. Geballe, and G. W. Hull, Jr., "Physical Properties of SmB$_6$," *Phys. Rev. B* **3**, 615 (1971).

[24] A. Jayaraman, V. Narayanamurti, E. Bucher, and R. G. Maines, "Continuous and Discontinuous Semiconductor-Metal Transition in Samarium Monochalcolenides Under Pressure," *Phys. Rev. Lett.* **25**, 1430 (1970).

[25] A. Sousanis, P. F. Smet, and D. Poelman, "Samarium monosulfide (SmS): Reviewing properties and applications," *Materials* **10**, 953 (2017). DOI: https://doi.org/10.3390/ma10080953

[26] M. B. Maple and D. Wohlleben, "Nonmagnetic 4f shell in the high-pressure phase of SmS," *Phys. Rev. Lett.* **27**, 511 (1971). http://dx.doi.org/10.1103/physrevlett.27.511

[27] M. B. Maple and D. Wohlleben, "Demagnetization of rare earth ions in metals due to valence fluctuations," *AIP Conference Proceedings* **18,** 447 (1974). http://dx.doi.org/10.1063/1.3141757

[28] Paul J. Robinson, Julen Munarriz, Michael E. Valentine, Austin Granmoe, Natalia Drichko, Juan R. Chamorro, Priscila F. Rosa, Tyrel M. McQueen, and Anastassia N. Alexandrova, "Dynamical Bonding Driving Mixed Valency in a Metal Boride," *Angewandte Chemie Int. Ed.* **59**, 10996 (2020). DOI: 10.1002/anie.202000945

[29] Priscila F. S. Rosa and Zachary Fisk, "Bulk and surface properties of $SmB_6$," arXiv:2007.09137v1 [cond-mat.str-el] 17 Jul 2020.  https://arxiv.org/pdf/2007.09137.

[30] M. Dzero, J. Xia, V. Galitski, and P. Coleman, "Topological Kondo insulators," *Annual Rev. Cond. Matt. Phys.* **7,** 249 (2016). DOI: https://doi.org/10.1146/annurev-conmatphys-031214-014749

[31] S. Wolgast, C. Kurdak, K. Sun, J. W. Allen, D.-J. Kim, and Z. Fisk, "Low-temperature surface conduction in the Kondo insulator $SmB_6$," *Phys. Rev. B* **88**, 180405 (R) (2013). DOI: 10.1103/PhysRevB.88.180405

[32] D. J. Kim, S. Thomas, T. Grant, J. Botimer, Z. Fisk, and J. Xia, "Surface Hall effect and nonlocal transport in $SmB_6$: Evidence for surface conduction," *Sci. Rep.* **3**, 3150 (2013). DOI: 10.1038/srep03150

[33] X. Zhang, N. P. Butch, P. Syers, S. Ziemak, R. L. Greene, and J. Paglione, "Hybridization, correlation, and in-gap states in the Kondo insulator $SmB_6$," *Phys. Rev. X* **3**, 011011 (2013). DOI: 10.1103/PhysRevX.3.011011

[34] E. S. Reich, "Hopes surface for exotic insulator," *Nature* **492**, 165 (2012).  DOI: https://doi.org/10.1038/492165a

[35] D. J. Kim, J. Xia, and Z. Fisk, "Topological surface state in the Kondo insulator samarium hexaboride," *Nature Matls.* **13,** 466 (2014). DOI: https://doi.org/10.1038/nmat3913

[36] G. Li, Z. Xiang, F. Yu, T. Asaba, B. Lawson, P. Cai, C. Tinsman, A. Berkley, S. Wolgast, Y. S. Eo, D-J. Kim, C. Kurdak, J. W. Allen, K. Sun, X. H. Chen, Y. Y. Wang, Z. Fisk, and Lu Li, "Two-dimensional Fermi surfaces in Kondo insulator SmB$_6$," *Science* **346**, 1208 (2014). DOI: https://doi.org/10.1126/science.1250366

[37] F. Chen, C. Shang, Z. Jin, D. Zhao, Y. P. Wu, Z. J. Xiang, Z. C. Xia, A. F. Wang, X. G. Luo, T. Wu, and X. H. Chen, "Magnetoresistance evidence of a surface state and a field-dependent insulating state in the Kondo insulator SmB$_6$," *Phys. Rev. B* **91**, 205133 (2015). DOI: https://doi.org/10.1103/PhysRevB.91.205133

[38] J. W. Allen, Forward, special issue of Philosophical Magazine on: "Topological correlated insulators and SmB$_6$," *Phil. Mag.* **96**, 3227 (2016).
DOI: 10.1080/14786435.2016.1247995

[39] J. Kim, C. Jang, X. Wang, J. Paglione, S. Hong, J. Lee, H. Choi, and D. Kim, "Electrical detection of the surface spin polarization of the candidate topological Kondo insulator SmB$_6$," *Phys. Rev. B* **99**, 245148 (2019) DOI: 10.1103/PhysRevB.99.245148

[40] Y. Ando, "Topological insulator materials," *J. Phys. Soc. Jpn.* **82**, 102001 (2013). DOI: https://doi.org/10.7566/JPSJ.82.102001

[41] J. G. Ramírez, A. C. Basaran, J. de la Venta, J. Pereiro, and I. K Schuller, "Magnetic field modulated microwave spectroscopy across phase transitions and the search for new superconductors," *Reports Prog. Phys.* **77**, 093902 (2014). DOI: https://doi.org/10.1088/0034-4885/77/9/093902

[42] B. Yang, M. Uphoff, Y-Q. Zhang, J. Reichert, A. P. Seitsonen, A. Bauer, C. Pfleiderer, and J. V. Barth, "Atomistic investigation of surface characteristics and electronic features at high-purity FeSi(110) presenting interfacial metallicity," *Proc. Natl. Acad. Sci. U.S.A.* **118**, e2021203118 (2021).
DOI: https://doi.org/10.1073/pnas.2021203118

[43] A. Kebede, M. C. Aronson, C. M. Buford, P. C. Canfield, J. H. Cho, B. R. Coles, J. C. Cooley, J. Y. Coulter, Z. Fisk, J. D. Goettee, W. L. Hults, A. Lacerda, T. D. McLendon, P. Tiwari, and J. L. Smith, "Studies of the correlated electron system SmB$_6$," *Physica B Condens. Matter* **223 & 224**, 256 (1996).
DOI: https://doi.org/10.1016/0921-4526(96)00092-0

[44] Yun Suk Eo, Alexa Rakoski, Juniar Lucien, Dmitri Mihaliov, Cağlıyan Kurdak, Priscila F. S. Rosa, and Zachary Fisk, "Transport gap in SmB$_6$ protected against disorder," *Proc. Natl. Acad. Sci. U.S.A.* **116**, 12638 (2019).
https://www.pnas.org/content/116/26/12638.

[45] J.-F. Lin, A. J. Campbell, D. L. Heinz, and G. Shen, "Static compression of iron-silicon alloys: Implications for silicon in the earth's core," *J. Geophys. Research: Solid Earth* **108**, 2045 (2003). DOI: https://doi.org/10.1029/2002JB001978

[46] R. A. Fischer, A. J. Campbell, D. M. Reaman, N. A. Miller, D. L. Heinz, P. Dera, and V. B. Prakapenka, "Phase relations in the Fe-FeSi system at high pressures and temperatures," *Earth and Planetary Science Letters* **373**, 54 (2013).
DOI: https://doi.org/10.1016/j.epsl.2013.04.035

[47] Z. Fisk, P. C. Canfield, J. D. Thompson, and M. F. Hundley, *J. Alloys Compounds* **181**, 369 (1992)

[48] E. Bauer, S. Bocelli, R. Hauser, F. Marabelli, and R. Spolenak, "Stoichiometric effects on the optical spectra and pressure response of $Fe_{1-x}Mn_xSi$," *Physica B: Condens. Matter* **230**, 794 (1997). DOI: https://doi.org/10.1016/S0921-4526(96)00842-3

[49] J. Beille, M. B. Maple, J. Wittig, Z. Fisk, and L. E. DeLong, "Suppression of the energy gap in $SmB_6$ under pressure," *Phys. Rev. B* **28**, 7397 (1983). DOI: https://doi.org/10.1103/PhysRevB.28.7397

[50] J. C. Cooley, M. C. Aronson, Z. Fisk, and P. C. Canfield, "$SmB_6$: Kondo insulator or exotic metal?" *Phys. Rev. Lett.*, **74**, 1629 (1995).
DOI: https://doi.org/10.1103/PhysRevLett.74.1629

[51] S. Gabáni, E. Bauer, S. Berger, K. Flachbart, Y. Paderno, Ch. Paul, V. Pavlík, and N. Shitsevalova, "Pressure-induced Fermi-liquid behavior in the Kondo insulator $SmB_6$: Possible transition through a quantum critical point," *Phys. Rev. B* **67**, 172406 (2003). DOI: https://doi.org/10.1103/PhysRevB.67.172406

[52] Y. Zhou, Q. Wua, P. F.S. Rosa, R. Yu, J. Guo, W. Yi, S. Zhang, Z. Wang, H. Wang, S. Cai, K. Yang, A. Li, Z. Jiang, S. Zhang, X. Wei, Y. Huang, P. Sun, Y-f. Yang, Z. Fisk, Q. Si, Z. Zhao, and L. Sun, "Quantum phase transition and destruction of Kondo effect in pressurized $SmB_6$," *Sci. Bull.* **62**, 1439 (2017).
https://doi.org/10.1016/j.scib.2017.10.008

[53] Y. Zhou, P. F. S. Rosa, J. Guo, S. Cai, R. Yu, S. Jiang, K. Yang, A. Li, Q. Si, Q. Wu, Z. Fisk, and L. Sun, "Hall-coefficient diagnostics of the surface state in pressurized $SmB_6$," *Phys. Rev. B* **101**, 125116 (2020). DOI: 10.1103/PhysRevB

[54] D. J. Campbell, Z. E. Brubaker, C. Roncaioli, P. Saraf, Y. Xiao, P. Chow, C. Kenney-Benson, D. Popov, R. J. Zieve, J. R. Jeffries, and J. Paglione, "Pressure-driven valence increase and metallization in the Kondo insulator $Ce_3Bi_4Pt_3$," *Phys. Rev. B* **100**, 235133 (2019). DOI: https://doi.org/10.1103/PhysRevB.100.235133

[55] Matthias Pickem, Emanuele Maggio and Jan M. Tomczak, "Resistivity saturation in Kondo insulators," *Communications Physics* **4**, 226 (2021).
https://doi.org/10.1038/s42005-021-00723-z

[56] P. Hlawenka, K. Siemensmeyer, E. Weschke, A. Varykhalov, J. Sánchez-Barriga, N. Y. Shitsevalova, A. V. Dukhnenko, V. B. Filipov, S. Gabáni, K. Flachbart, O. Rader, and E. D. L. Rienks, "Samarium hexaboride is a trivial surface conductor," *Nature Commun.* **9**, 517 (2018). DOI: https://doi.org/10.1038/s41467-018-02908-7

[57] S. M. Thomas, X. Ding, F. Ronning, V. Zapf, J. D. Thompson, Z. Fisk, J. Xia, and P. F. S. Rosa, "Quantum oscillations in flux-grown SmB$_6$ with embedded aluminum," *Phys. Rev. Lett.* **122**, 166401 (2019). DOI: https://doi.org/10.1103/PhysRevLett.122.166401